\documentclass[10pt, twocolumn, nofootinbib,superscriptaddress]{revtex4-1}
\usepackage{amsmath,amssymb,amsfonts}
\usepackage{algorithmic}
\usepackage{graphicx}
\usepackage{textcomp}
\usepackage{xcolor}
\usepackage{ragged2e}
\usepackage{booktabs, makecell, tabularx}

\usepackage{gensymb}
\makeatletter
    \renewcommand\@make@capt@title[2]{%
     \@ifx@empty\float@link{\@firstofone}{\expandafter\href\expandafter{\float@link}}%
      {\textbf{#1}}\@caption@fignum@sep#2\quad}%
\makeatother
\makeatletter 
\renewcommand{\fnum@figure}{\textbf{Fig.~\thefigure}} 
\makeatother

\usepackage{xcolor}
\newcommand{\beginsupplement}{%
        \setcounter{table}{0}
        \renewcommand{\thetable}{S\arabic{table}}%
        \setcounter{figure}{0}
        \renewcommand{\thefigure}{S\arabic{figure}}%
     }
\def\BibTeX{{\rm B\kern-.05em{\sc i\kern-.025em b}\kern-.08em
    T\kern-.1667em\lower.7ex\hbox{E}\kern-.125emX}}

\begin{document}

\author{Gaojian Liu}
\thanks{These authors contributed equally}
\affiliation{Nonlinear Nanophotonics Group, MESA+ Institute of Nanotechnology,\\
University of Twente, Enschede, Netherlands}
\affiliation{China Academy of Space Technology (Xi'an), Xi'an, China}
\author{Kaixuan Ye}
\thanks{These authors contributed equally}
\affiliation{Nonlinear Nanophotonics Group, MESA+ Institute of Nanotechnology,\\
University of Twente, Enschede, Netherlands}
\author{Okky Daulay}
\affiliation{Nonlinear Nanophotonics Group, MESA+ Institute of Nanotechnology,\\
University of Twente, Enschede, Netherlands}
\author{Qinggui~Tan}
\affiliation{China Academy of Space Technology (Xi'an), Xi'an, China}
\author{Hongxi~Yu}
\affiliation{China Academy of Space Technology (Xi'an), Xi'an, China}
\author{David Marpaung}
\email{david.marpaung@utwente.nl}
\affiliation{Nonlinear Nanophotonics Group, MESA+ Institute of Nanotechnology,\\
University of Twente, Enschede, Netherlands}

\date{\today}

\title{Linearized Integrated Microwave Photonic Circuit for Filtering and Phase Shifting}

\begin{abstract}
Photonic integration, advanced functionality, reconfigurability, and high RF performance are key features in integrated microwave photonic systems that are still difficult to achieve simultaneously. In this work, we demonstrate an integrated microwave photonic circuit that can be reconfigured for two distinct RF functions, namely, a tunable notch filter and a phase shifter. We achieved $>$50~dB high-extinction notch filtering over 6-16 GHz and 2$\pi$ continuously tunable phase shifting over 12-20 GHz frequencies.  At the same time, we implemented an on-chip linearization technique to achieve a spurious-free dynamic range of more than  120~$\rm{dB}\cdot \rm{Hz}^{4/5}$ for both functions. Our work combines multi-functionality and linearization in one photonic integrated circuit, and paves the way to reconfigurable  RF photonic front-ends with very high performance. 
\end{abstract}
\maketitle

\section*{Introduction}
The development of cognitive and intelligent radio frequency (RF) systems calls for RF front-ends with programmable multiple functions \cite{Baylis2014solving,Islam2020diode}. Microwave photonic (MWP) circuits with high bandwidth and  flexible reconfigurability can play an important role in those modern RF systems \cite{capmany2007microwave, marpaung2019integrated,Zhu2020Broadband}. A number of approaches have been proposed to achieve programmable RF photonic circuits. Functionalities including filtering \cite{fandino2017monolithic,liu2020integrated,daulay2020chip, Chen2023Multiband,tao2021hybrid}, phase shifting \cite{chew2022Inline,porzi2018photonic,McKay2019Brillouin,lin2020high,chew2019integrated}, and beamforming \cite{zhu2020silicon,Romero2022high,Lin2022low} have been realized in application-specific photonic integrated circuits (PIC) that can achieve high performance. Another approach is to create general-purpose integrated MWP circuits that can be configured for a high number of functionalities. These circuits have been demonstrated in the form of  waveguide mesh or micro-disk resonator array \cite{zhuang2015programmable,perez2017multipurpose,bogaerts2020,zhang2020photonic}. Finally, cascaded MWP systems that can simultaneously perform notch filtering and phase shifting have also been demonstrated in \cite{Nguyen2016,Xue2012}.

To be relevant for real applications, MWP circuits need to be more than just programmable. Sufficient RF performance such as low noise figure and high spurious-free dynamic range (SFDR) is also required so that they can be placed at the beginning of the RF front-end chain, where the broad bandwidth is most needed.   



However, to date, most demonstrated integrated MWP systems are plagued by poor SFDR \cite{tao2021hybrid,fandino2017monolithic,Yao2022silicon,LiuY2022tunable}. There is an opportunity to circumvent this problem by implementing linearization techniques \cite{wu2019multi,Wang2022high,liu2021integrated}. But only very recently can the RF photonic functionality and linearization be simultaneously achieved in the same system, for example in fiber-based phase shifters \cite{bai2021large}, or in a chip-based programmable filter \cite{daulay2022ultrahigh}. Extending this concept to the multi-functional circuit will be important for the MWP field. 


\begin{figure*}[t!]
\centering
\includegraphics[width=\linewidth]{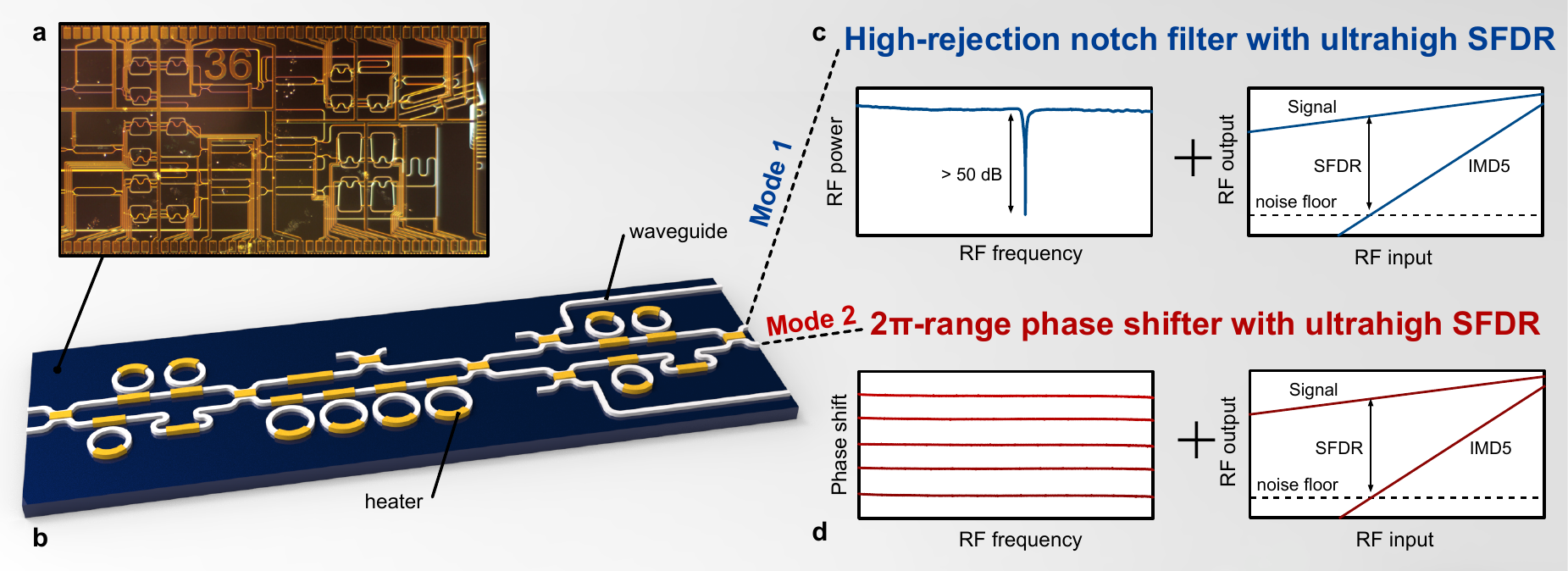}
\caption{\textbf{Concept of the reconfigurable integrated MWP circuit with enhanced dynamic range.} \textbf{(a)} The micrograph of the fabricated chip. \textbf{(b)} Schematic of the reconfigurable integrated MWP circuit.  \textbf{(c-d)} The illustration of the achieved functionalities.}
\label{fig1}
\end{figure*}

In this work, we present a reconfigurable integrated MWP circuit that can perform two distinct functionalities, i.e., a tunable notch filter and a 2$\pi$-range phase shifter, while maintaining an ultrahigh SFDR. The tunable notch filter function can achieve rejection of more than 50 dB with a 3-dB bandwidth of 315~MHz from 6 to 16~GHz. In the meantime, the tunable phase shifter function can realize 2$\pi$ continuous tuning range from 12 to 20~GHz. Moreover, both functions can provide a SFDR larger than 120~$\rm{dB}\cdot\rm{Hz}^{4/5}$. This work shows the important first step towards high-performance integrated MWP subsystems with a large number of RF functions.


\section*{Results}
\subsection*{Reconfigurable Integrated MWP Circuit}

The concept of our reconfigurable integrated MWP circuit is illustrated in Fig.~\ref{fig1}. It can serve as a notch filter or a phase shifter, with different configurations. By manipulating the phases and amplitudes of multi-order optical sidebands independently, both functionalities can also exhibit ultrahigh dynamic range. 

The multi-functional integrated MWP circuit is fabricated with the Si$_3$N$_4$ TriPleX process \cite{roeloffzen2018low,worhoff2015triplex}. Fig.~\ref{fig1}(a) shows the micrograph of the chip. It consists of two spectral de-interleavers, a tunable attennuator, a phase shifter, and an array of all-pass ring resonators, as illustrated in Fig.~\ref{fig1}(b). The spectral de-interleavers are implemented with an asymmetric Mach Zehnder interferometer (aMZI) loaded with 3 ring resonators \cite{luo2010high}, exhibiting flat-top complementary filter response with free spectral range (FSR) of 160 GHz \cite{liu2021integrated}. The array of all-pass ring resonators (FSR of 50~GHz) between the two de-interleavers are with tunable coupling ratio and round trip phase that can be controlled via thermo-optic tuning. By applying different voltages to the microheaters, this chip can be configured as a notch filter or a phase shifter while keeping an ultrahigh SFDR, as indicated in Fig.~\ref{fig1}(c) and (d).

The signal processing of both functionalities follows a similar procedure. First, the de-interleaver separates the input optical spectrum into two paths. The attenuator and phase shifter in the upper path change the relative phase and amplitude between the two paths for wideband spectrum shaping. In the lower path, an array of all-pass ring resonators performs narrowband phase and amplitude tailoring with high precision. The signals in the two paths are then recombined and directly coupled out of the chip, or coupled into the second de-interleaver for further processing. By tailoring the amplitudes and phases of the optical carrier and multi-order sidebands in such a manner, the $3^{rd}$-order intermodulation distortion (IMD3) from beating products between different optical sidebands can destructively interfere with each other, leading to an ultrahigh dynamic range. 

\begin{figure*}[ht!]
	\centering
	\includegraphics[width=\textwidth]{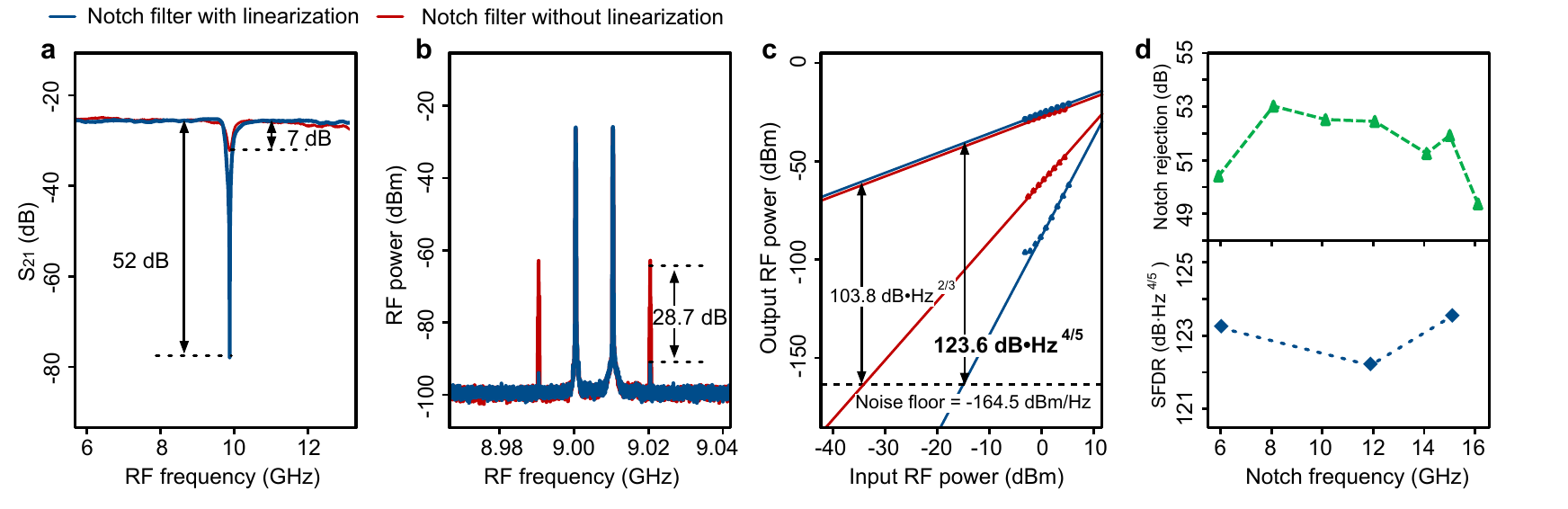}
	\caption{\textbf{High-rejection notch filter with ultrahigh dynamic range.} \textbf{(a)} Measured linearized RF notch response with rejection of 52 dB compared with the bench-marked SSB RF notch filter. \textbf{(b)} The measured two-tone RF spectra at the output of the photodetector for the linearized RF notch filter (blue) and single-sideband (SSB) RF notch filter	without linearization (red) with 28.7 dB reduction of IMD3 power. \textbf{(c)} The measured spurious-free dynamic range (SFDR) of the proposed linearized notch filter and SSB notch filter without linearization at RF frequency of 9 GHz. The proposed linearized filter has a record-high SFDR of 123.6~$\rm{dB}\cdot \rm{Hz}^{4/5}$. \textbf{(d)} The measured rejection of the linearized notch filter and the SFDR at 9 GHz with the tuning of the notch frequency.}
	\label{fig2}
\end{figure*}

\subsection*{High-rejection notch filter with ultrahigh SFDR}
\label{susec:two}

We first configure this multi-functional MWP circuit into a high-rejection notch filter with enhanced SFDR. A phase-modulated optical signal is sent to the programmable photonic chip for spectral shaping. Because of the nonlinearity of the phase modulator, the modulated optical signal consists of multi-order sidebands, among which the $\pm~1^{st}$-order sidebands have a $\pi$ phase difference. We use the spectral de-interleaver to first separate the optical carrier and lower sidebands from the upper sidebands. Then, the lower sidebands and optical carrier are attenuated to generate the asymmetric double sideband (aDSB) spectrum and simultaneously satisfying the linearization condition \cite{daulay2022ultrahigh}. Meanwhile, an under-coupled ring resonator imposes a shallow notch at the upper sideband to meet equal amplitude and anti-phase conditions at the notch frequency. After that the signal is recombined, and a high rejection MWP notch filter based on RF interference is created \cite{marpaung2013si}. Because of the destructive interference between different IMD3 components, the linearity of the notch filter can be significantly improved (See Supplementary Note for more experimental details). 

The performance of the linearized MWP notch filter is shown in Fig.~\ref{fig2}. To highlight the advantages of our filter, we compared it with a standard single sideband (SSB) modulated MWP notch filter using the same ring response without linearization. Our demonstrated notch filter has a notch rejection of 52 dB and a 3-dB bandwidth of 315 MHz (at the center frequency of 12 GHz), as shown in Fig.~\ref{fig2}(a). We also performed the two-tone test to evaluate the linearity of the demonstrated notch filter. The results shown in Fig.~\ref{fig2}(b) exhibit an IMD3 suppression of 28.7 dB when the two-tone RF frequency is at 9~GHz with a spacing of 10~MHz. The large suppression of the IMD3 makes the $5^{th}$-order nonlinearity dominate, leading to SFDR improvement from 103.8 $\rm{dB}\cdot \rm{Hz}^{2/3}$ to 123.6 $\rm{dB}\cdot \rm{Hz}^{4/5}$ with a noise floor of -164.5 dBm/Hz, as observed in Fig.~\ref{fig2}~(c). Moreover, the filter keeps high rejection and large SFDR during the tuning of the notch frequency as shown in Fig.~\ref{fig2}~(d) The filter shows a link gain of -26.3 dB and a noise figure (NF) of 35.8 dB. This moderate gain and NF is because of the partially destructive interference between two $1^{st}$-order sidebands (See Supplementary Note for extended measurement results).

\begin{figure*}[ht!]
\centering
\includegraphics[width=\textwidth]{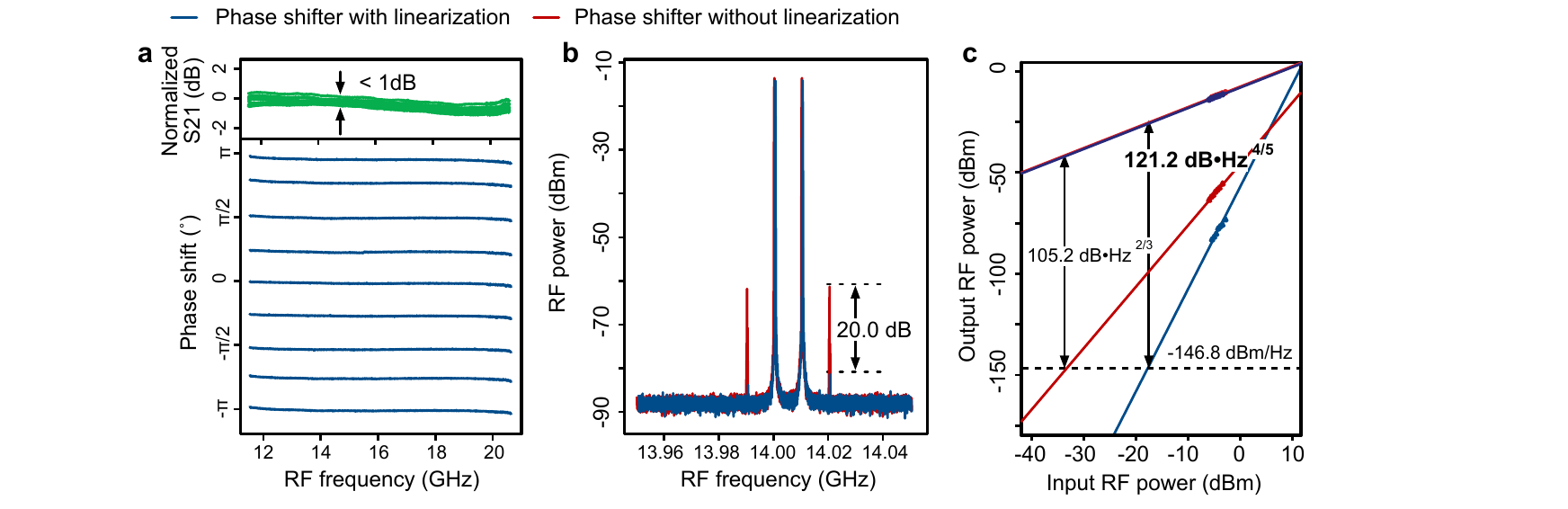}
\caption{\textbf{2$\pi$-range tunable phase shifter with ultrahigh dynamic range.} \textbf{(a)} The phase tuning of the proposed phase shifter over 2$\pi$. The phase response is flat in the range of 12-20 GHz. \textbf{(b)} The measured two-tone RF spectra at the output of the photodetector for the linearized RF phase shifter (blue) and bench-marked RF phase shifter without linearization (red) with 20.0 dB improvement of fundamental to IMD3 ratio. \textbf{(c)} The measured spurious-free dynamic range (SFDR) of the proposed linearized phase shifter and bench-marked phase shifter without linearization at RF frequency of 14 GHz. The proposed phase shifter exhibits a record-high SFDR of 121.2~$\rm{dB}\cdot \rm{Hz}^{4/5}$.}
\label{fig3}
\end{figure*}

\subsection*{2$\pi$-range phase shifter with ultrahigh SFDR}
\label{susec:three}

The chip can also be reconfigured as a 2$\pi$-range tunable phase shifter while maintaining the ultrahigh dynamic range. To realize the phase shifting function, we first use  an external band pass filter (BPF) to remove the lower sideband of the phase-modulated optical signal. We also attenuate the optical carrier with the BPF by positioning the optical carrier at the transition band for the linearization purpose. The preprocessed single-sideband (SSB) optical signal is then coupled into the chip. We separate the optical carrier and the $+1^{st}$-order sideband from the $+2^{nd}$-order sideband with the spectral de-interleaver. By applying two over-coupled ring resonators at the  optical carrier, 2$\pi$-range phase shift is introduced with negligible amplitude variation. The optical phase shift is then converted to the RF phase shift when the optical carrier beats with the $+1^{st}$-order sideband at the photodetector (See the Supplementary Note for the experimental setup). To suppress the IMD3 and improve the linearity of the phase shifter, both amplitude and phase conditions between the optical carrier and the $+2^{nd}$-order sideband need to be satisfied (See the Supplementary Note for the linearization method). On the one hand, we adjust the power ratio between the optical carrier and $+2^ {nd}$-order sideband by controlling the passband location of the external BPF and the coupling coefficients of the ring resonators. On the other hand, we tune the phase of the $+2^{nd}$-order sideband with an optical phase shifter to satisfy the phase condition, while the MWP phase shifter tuning over the 2$\pi$ range.

The performance of the demonstrated phase shifter is presented in Fig.~\ref{fig3}. We can realize continuous phase tuning over 2$\pi$ range by adjusting the resonance frequencies of the two over-coupled rings. Because of the relatively flat stopband response of the two cascaded ring resonators, the amplitude fluctuation of the phase shifter is less than 1~dB over the whole 2$\pi$ range, as shown in Fig.~\ref{fig3} (a). To evaluate the linearity of the demonstrated phase shifter, we also conducted a two-tone test and compared the results with a phase shifter without linearization for the same phase shift (0.6~$\pi$ in this case). As shown in Fig.~\ref{fig3}(b), the linearized phase shifter exhibits an improved fundamental to IMD3 ratio of 20~dB. As a result, the SFDR of the phase shifter increases from 105.2 $\rm{dB}\cdot \rm{Hz}^{2/3}$  to 121.2 $\rm{dB}\cdot \rm{Hz}^{4/5}$, which are shown in Fig.~\ref{fig3}(c). The noise floor of both phase shifters is \mbox{-146.8~dBm/Hz}. The phase shifter exhibits a link gain of -5.8 dB and NF of 33 dB. The NF of the proposed phase shifter is mainly deteriorated by the phase noise to intensity noise conversion of the laser when the optical carrier is attenuated by the external BPF. 

\section*{Discussion and Conclusion}
In this work, we demonstrated a multi-functional integrated MWP circuit that can be reconfigured between a high-rejection notch filter and a 2$\pi$-range phase shifter. By vectorially manipulating the multi-order optical sidebands, ultrahigh dynamic range over 120 $\rm{dB}\cdot \rm{Hz}^{4/5}$ is achieved in both functions. The array of microring resonators and two spectral de-interleavers lead to a versatile circuit, and could potentially unlock even more functionalities apart from filter and phase shifting, for example, true time delay. Moreover, the possibility to realize multiple different functions in one circuit also opens the path to the cascaded integrated MWP system, which would be a major leap towards integrated RF photonic front-ends that can be directly applied in real RF environments. 

Currently, the reconfigurable MWP circuit exhibits relatively low link gain and high noise figure. In principle, the link gain can be improved by increasing the optical power to a high power-handling modulator. The strategy for noise figure reduction, on the other hand, is more intricate. The use of the phase modulator in our experiments prevents NF reduction techniques such as low-biasing a Mach-Zehnder intensity modulator \cite{daulay2022ultrahigh} to be employed. Emulation of such technique using carrier processing by ring resonator has been previously considered \cite{daulay2020chip} but the power builds up in the high-Q ring used for carrier processing, preventing the effective NF reduction. We believe that achieving simultaneous high link gain and low NF together with the features demonstrated in this work will require an entirely new topology potentially using a more complex interferometric modulator (for example dual-drive or dual-parallel MZM) with on-chip linearization. Several recent works have shown encouraging results in this direction \cite{feng2022ultrahighlinearity,moran2023linearization}.

\section*{Author Contribution}
G.L. and K.Y. contributed equally in this work. D.M. and G.L. developed the concept and proposed the physical system. O.D. designed the photonic circuits, G.L. and K.Y. developed and performed numerical simulations. G.L. and K.Y. performed the experiments with input from O.D.. D.M., G.L. and K.Y. wrote the manuscript with input from Q.T., H.Y. D.M. led and supervised the entire project.
\label{sec:four}

\section*{Funding Information}

This project is funded by the European Research Council Consolidator Grant (101043229 TRIFFIC) and Nederlandse Organisatie voor Wetenschappelijk Onderzoek (NWO) projects (740.018.021 and 15702).



\bibliographystyle{IEEEtran}
\bibliography{library}

\newpage
\onecolumngrid
\beginsupplement

\newpage

\section*{Supplementary Note: Experimental details}
In the supplementary note, we present the experimental details of both the notch filter and the phase shifter. We start from the experimental setups and signal flows. Then, we derive the linearization conditions by manipulating the amplitude and phase of the optical carrier and the multi-order sidebands. Finally, we show extended measurement results of both experiments.

\subsection*{Experimental Setups}
\begin{figure*}[ht!]
\centering
\includegraphics{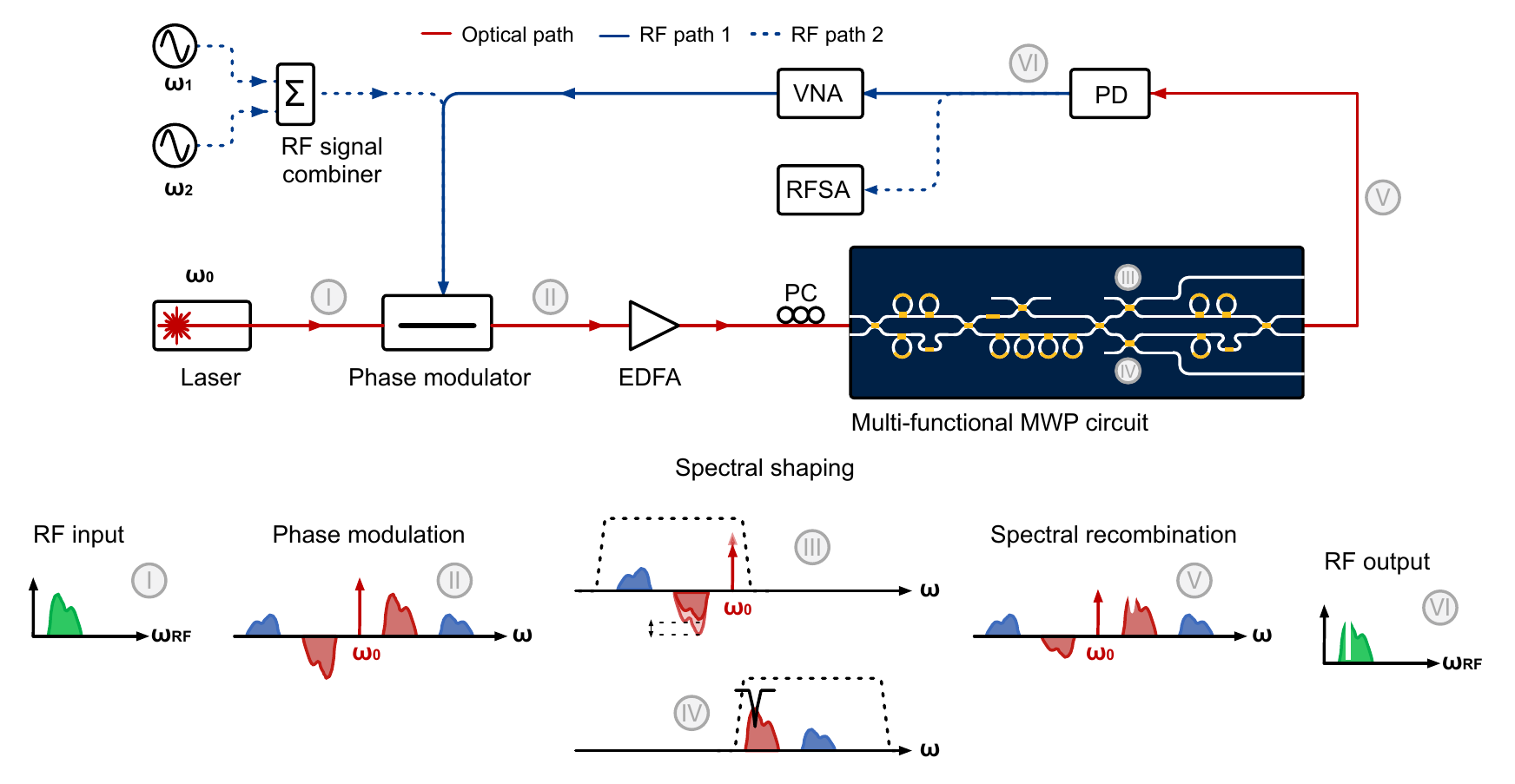}
\caption{Experiment setup and signal flow for demonstrating linearized notch filter.}
\label{figs1}
\end{figure*}

The experimental setup for the notch filter is shown in Fig.~\ref{figs1}. Light from the low relative-intensity noise (RIN) CW laser (Pure Photonics PPCL550) is set at 1550~nm and 18~dBm. The light is modulated with RF signal via a phase modulator (EOSpace, 20 GHz). Because of the nonlinearity of the phase modulator, higher-order sidebands would be generated along with the $1^{st}$-order sidebands. The signal is amplified by a low noise erbium-doped fiber amplifier (EDFA, Amonics) before injected into the chip for signal processing. The on-chip deinterleaver separates the signal into two path, one containing the optical carrier and lower-sidebands and the other one containing upper sidebands. In the subsequent processing, a ring resonator is set at the under-coupling state and creates a notch response at the $1^{st}$-order upper sideband. In the meantime, the $1^{st}$-order lower sideband and optical carrier is suppressed with the tunable attenuator. After that, signals at two paths are recombined and sent to the photo detector (EMCORE 20GHz) to be converted to the RF domain. Since the 1st-order upper and lower sidebands at the notch frequency are equal in amplitude but have anti-phase. A very deep notch can be formed at that frequency because of the destructive interference in the RF domain. The filter response is measured with the VNA (keysight P9375A). For the SFDR measurement, the phase modulator is driven by a two-tone signal with a spacing of 10 MHz generated by RF signal generators (Wiltron 69147A and Rohde-Schwarz SMP02), and the converted RF signal is measured with an RF spectrum analyzer (Keysight N9000B).

\begin{figure*}[ht!]
\centering
\includegraphics{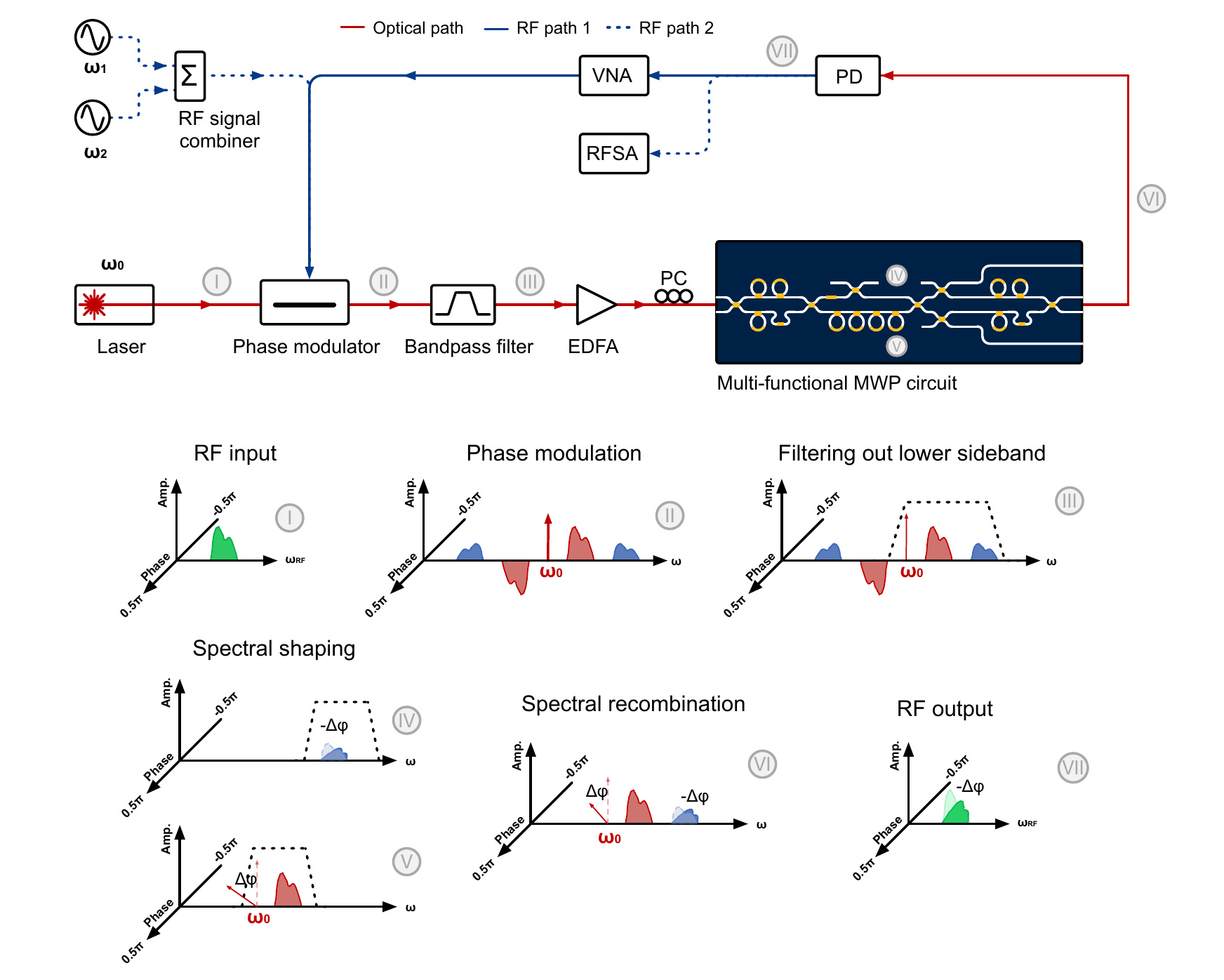}
\caption{Experiment setup and signal flow for demonstrating linearized phase shifter.}
\label{figs2}
\end{figure*}

The setup and signal flow of the phase shifter experiment is shown in Fig.~\ref{figs2}. Unlike the setup of the notch filter experiment in Fig.~\ref{figs1}, we use an external bandpass filter (EXFO XTM-50) to filter out the low sidebands and also introduce 9~dB attenuation to the optical carrier for linearization (as discussed later). Then, the signal is amplified by a low-noise EDFA (Amonics) and injected into the chip. We further use the on-chip deinterleaver to spatially separate the 2$^{\rm nd}$-order upper sideband from the optical carrier and the 1$^{\rm st}$-order upper sideband into two paths. For the path with the optical carrier, we apply two cascaded all-pass ring resonators at the over-coupled state to introduce phase shift to the optical carrier. On the other hand, for the path with the 2$^{\rm nd}$-order upper sideband, we use the thermal-optic phase shifter to create the opposite phase to the 2$^{\rm nd}$-order upper sideband and use the tunable attenuator to adjust the power ratio between 
 the optical carrier and $2^{nd}$-order sideband for the best IMD3 suppression. Finally, the signals at two paths recombine and are converted back to the RF domain with the photo detector (APIC, 40 GHz), and the converted RF signal is measured with a VNA (keysight P9375A). The settings of the SFDR measurement of the phase shifter is the same as the notch filter.

\subsection*{Linearization Method}

Phase-modulated optical spectrum consists of multi-order optical sidebands which would generate intermodulation distortions in RF domain after photodetection. We manipulate the phase and amplitude of these multi-order optical sidebands to achieve selective destructive interference among the IMD3 terms to improve the linearity of the integrated MWP systems. As the linearization condition of the notch filter has been extensively explained in \cite{daulay2022ultrahigh}, here we only derive the linearization condition of the phase shifter. 

We adopt a two-tone analysis to investigate the main contributors of IMD3 terms and the requirements to suppress them. When the optical carrier is modulated by a two-tone RF signal at frequencies $\rm \omega_1$ and $\rm \omega_2$ with a phase modulator, the optical field after the PM can be expressed as:

\begin{equation}\label{eq:PhaseinField}
    E_{\rm p}(t) = \sqrt{P_{in}}\ e^{j\omega_c t}\sum_{n=-\infty}^{+\infty}\sum_{k=-\infty}^{+\infty}J_n(m)J_k(m)e^{j(n\omega_1 + k\omega_2)t}
\end{equation}

Where $P_{in}$ is the input power, $\omega_c$ is the optical carrier frequency, $J_n$ is the $n^{th}$-order Bessel function of the first kind, and $m$ is the modulation index, which can be expressed as: $m = \pi \cdot V_{RF}/V_{\pi}$. 

To simplify the analysis, we assume that the two-tone RF signal at frequencies $\rm \omega_1$ and $\rm \omega_2$ are small signals, we only need to consider up to the 2$^{\rm nd}$-order sidebands. We assume the attenuation to the optical carrier and $2^{nd}$-order sideband is $\alpha_0$ and $\alpha_2$ respectively. Then the simplified optical field after the chip can be written as:

\begin{equation}\label{eq:phaseOutField}
E_{\mathrm{out}}=\sqrt{P_{i}} \cdot e^{j \omega_{c} t}\left\{\begin{array}{l}
\alpha_{0}\left[J_{0}^{2} e^{j \Delta \varphi_{0}}-J_{1}^{2} e^{j\left[\left(\omega_{1,2}-\omega_{2,1}\right) t+\Delta \varphi_{0}\right]}\right] \\
+\left[J_{0} J_{1} e^{j \omega_{1,2} t}-J_{1} J_{2} e^{j\left[\left(2 \omega_{1,2}-\omega_{2,1}\right) t\right]}\right] \\
+\alpha_{2} J_{0} J_{2} e^{j\left(2 \omega_{1,2} t+\Delta \varphi_{2}\right)}
\end{array}\right\}
\end{equation}

The photo-current of the RF signal detected from the processed optical spectrum can be expressed as

\begin{equation}\label{eq:photocurrent}
    \begin{aligned}
I_{P D}(t) &=R_{P D}\left|E_{p}(t)\right|^{2} \\
&=I_{1} \cos \omega_{1,2} t+I_{3} \cos \left(2 \omega_{1,2}-\omega_{2,1}\right) t
\end{aligned}
\end{equation}

where $R_{PD}$ is the responsivity of the photo detector, $I_1$ and $I_3$ are the amplitude coefficients for fundamental signal and IMD3 components, which can be written as

\begin{equation}
\begin{aligned}
&i_{\mathrm{RF}}(t) \propto \alpha_{0} J_{0}^{3} J_{1} \cdot \cos \left[\omega_{1,2} t-\Delta \varphi_{0}\right] \\
&i_{I M D 3}(t) \propto \alpha_{2} J_{0}^{2} J_{1} J_{2} \cdot \cos \left[\left(2 \omega_{1,2}-\omega_{1,2}\right) t+\Delta \varphi_{2}\right]-\alpha_{0}\left(J_{0}^{2} J_{1} J_{2}+J_{0} J_{1}^{3}\right) \cdot \cos \left[\left(2 \omega_{1,2}-\omega_{1,2}\right) t-\Delta \varphi_{0}\right]
\end{aligned}
\end{equation}

To minimize the IMD3 terms and maximize the fundamental RF signal, the attenuation and phase shift imposed to the optical carrier and $+2^{nd}$-order optical sideband should satisfy the condition:

\begin{equation}\label{eq:condition1}
\left\{\begin{array}{l}
\Delta \varphi_{0}=-\Delta \varphi_{2} \\
\alpha_{2} J_{0}^{2} J_{1} J_{2}-\alpha_{0}\left(J_{0}^{2} J_{1} J_{2}+J_{0} J_{1}^{3}\right)=0
\end{array}\right.
\end{equation}

If we take small signal approximation and apply Taylor expansion to the Bessel function of the first kind, then the condition in \eqref{eq:condition1} can be simplified as:

\begin{equation}\label{eq:condition2}
\left\{\begin{array}{l}
\Delta \varphi_{0}=-\Delta \varphi_{2} \\
\alpha_{2}m^3 - 3\alpha_{0}m^3=0
\end{array}\right.
\end{equation}

This condition means that the phase shift added to $+2^{nd}$-order optical sideband is the opposite of that of the optical carrier. Moreover, the attenuation imposed to the electrical field of the optical carrier is 3 times of that applied to the $+2^{nd}$-order optical sideband. This means that the power attenuation at optical carrier is 9.54 dB larger than the power attenuation of the $+2^{nd}$-order optical sideband, and the phase shift imposed to the RF signal is $-\Delta \varphi_0$.

\subsection*{Extended Measurement Results}

\begin{table}[ht]
\caption{\textbf{Measured Fundamental to IMD3 ratio and the SFDR of the notch filter at different two-tone RF frequencies.}}
\label{tab:notchfilter}
\begin{tabular}{c|c|c}
    \textbf{RF frequencies} &\textbf{Fundamental to IMD3 ratio } & \textbf{SFDR of the notch filter with linearization (without linearization)}\\
    \textbf{GHz} & \textbf{dB} & \textbf{dB$\cdot$Hz$^{4/5}$ (dB$\cdot$Hz$^{2/3}$)}\\
    \hline
    8  & 63.8  & 122.6 (102.8)\\
    9  & 64  & 122.3 (103.7)\\
    10  & 63.2  & 123.5 (102.5)\\
    16  & 61.6  & 122.6 (103.5)\\
\end{tabular}
\end{table}

To further characterize the performance of our proposed linearized RF notch filter, we extend our measurements of IMD3 suppression and SFDR when the two-tone test frequency and the notch filter frequency are tuned separately.
We first fix the notch frequency at 12 GHz and performed two-tone test at 8~GHz, 9~GHz, 10~GHz, and 16~GHz. The results of the IMD3 suppression and the SFDR at different frequencies are listed in Table \ref{tab:notchfilter}. The IMD3 terms are greatly suppressed and fundamental to IMD3 ratios of more than 61~dB can be achieved in all these two-tone frequencies. Moreover, SFDR of more than 122 dB$\cdot \rm Hz^{4/5}$ are observed with improvements around 20~dB compared with nonlinearized states.


We also measured SFDR at a specific frequency (in this case 9~GHz), when the notch response is tuned at 6~GHz, 12~GHz, and 15GHz. The results are listed in Table \ref{tab:differentNotch}. In all these frequencies the improved SFDR are more than 122 dB$\cdot \rm Hz^{4/5}$. These results further demonstrate that the proposed linearized notch filter has a high linearity and large SFDR in a wide frequency range.

\begin{table}[ht]
\caption{\textbf{Measured SFDR of the notch filter at different notch frequencies.}}
\label{tab:differentNotch}
\begin{tabular}{c|c|c|c}
    \textbf{Notch frequencies} & 6 GHz & 12 GHz & 15 GHz\\
    \hline
    \textbf{SFDR} & 123.2 dB$\cdot$Hz$^{4/5}$ & 122.3 dB$\cdot$Hz$^{4/5}$ &123.6 dB$\cdot$Hz$^{4/5}$\\
\end{tabular}
\end{table}



\begin{figure*}[ht!]
\centering
\includegraphics[width=\linewidth]{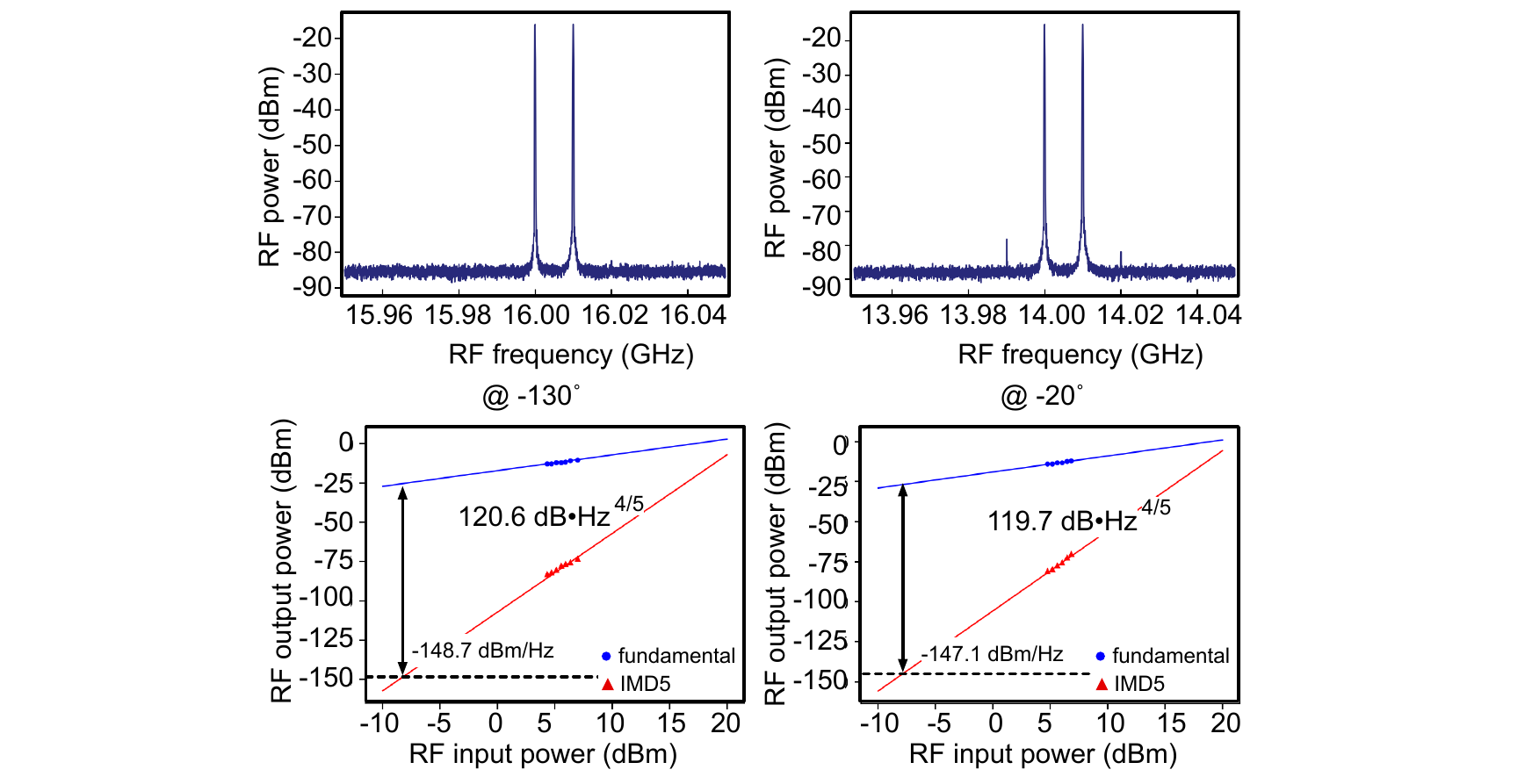}
\caption{IMD3 suppression and SFDR measurements of the phase shifter at various two-tone frequencies and phase shifts.}
\label{figs6}
\end{figure*}

To further characterize the performance of our proposed linearized RF phase shifter, we extend our measurements of IMD3 suppression and SFDR when the two-tone test frequency and the phase shifting are tuned. We performed two-tone test at 16~GHz, 14~GHz with  phase shifting of -130$^{\circ}$, -20$^{\circ}$  respectively. The results of the IMD3 suppression and SFDR are shown in Fig.~\ref{figs6}. It is clear that the IMD3 terms are greatly suppressed and the SFDRs keep enhanced close to 120 dB$\cdot \rm Hz^{4/5}$ in all combination of two-tone frequency and phase shifting.

\end{document}